\documentclass[%
superscriptaddress,
showpacs,
 amsmath,amssymb,
aps,
prl,
twocolumn,
]{revtex4-1}

\usepackage{array}
\usepackage{wrapfig}

\usepackage{graphicx}
\usepackage{dcolumn}
\usepackage{bm}
\usepackage{amsmath}
\usepackage{amssymb}
\usepackage{graphics}
\usepackage{epsfig}
\usepackage{natbib}
\usepackage{verbatim} 

\usepackage{color}

\usepackage{siunitx} 

\newcommand{\muSR}{$\mu$SR }

\begin{document}


\title{Coexistence of low moment magnetism and superconductivity in tetragonal FeS and suppression of $T_\mathrm{c}$ under pressure}

\author{S.~Holenstein}
\affiliation{Laboratory for Muon Spin Spectroscopy, Paul Scherrer
Institute, CH-5232 Villigen PSI, Switzerland}
\affiliation{Physik-Institut der Universit\"at Z\"urich,
Winterthurerstrasse 190, CH-8057 Z\"urich, Switzerland}
\author{U.~Pachmayr}
\affiliation{Department Chemie, Ludwig-Maximilians-Universit\"at M\"unchen,
Butenandtstr. 5-13 (D), 81377 M\"unchen, Germany}
\author{Z.~Guguchia}
\affiliation{Laboratory for Muon Spin Spectroscopy, Paul Scherrer
Institute, CH-5232 Villigen PSI, Switzerland}
\author{S.~Kamusella}
\affiliation{Institut f\"ur Festk\"orperphysik, TU Dresden, DE-01069
Dresden, Germany}
\author{R.~Khasanov}
\affiliation{Laboratory for Muon Spin Spectroscopy, Paul Scherrer
Institute, CH-5232 Villigen PSI, Switzerland}
\author{A.~Amato}
\affiliation{Laboratory for Muon Spin Spectroscopy, Paul Scherrer
Institute, CH-5232 Villigen PSI, Switzerland}
\author{C.~Baines}
\affiliation{Laboratory for Muon Spin Spectroscopy, Paul Scherrer
Institute, CH-5232 Villigen PSI, Switzerland}
\author{H.-H.~Klauss}
\affiliation{Institut f\"ur Festk\"orperphysik, TU Dresden, DE-01069
Dresden, Germany}
\author{E.~Morenzoni}
\affiliation{Laboratory for Muon Spin Spectroscopy, Paul Scherrer
Institute, CH-5232 Villigen PSI, Switzerland}
\affiliation{Physik-Institut der Universit\"at Z\"urich,
Winterthurerstrasse 190, CH-8057 Z\"urich, Switzerland}
\author{D.~Johrendt}
\affiliation{Department Chemie, Ludwig-Maximilians-Universit\"at M\"unchen,
Butenandtstr. 5-13 (D), 81377 M\"unchen, Germany}
\author{H.~Luetkens}
\email[Corresponding
author:~]{hubertus.luetkens@psi.ch}
\affiliation{Laboratory for Muon Spin Spectroscopy, Paul Scherrer
Institute, CH-5232 Villigen PSI, Switzerland}

\begin{abstract}
We report local probe ($\mu$SR) measurements on the recently discovered tetragonal FeS superconductor which has been predicted to be electronically very similar to superconducting FeSe. Most remarkably, we find that low moment ($10^{-2}-10^{-3}\mu_\mathrm{B}$) disordered magnetism with a transition temperature of $T_\mathrm{N}\approx 20$~K microscopically coexists with bulk superconductivity below $T_\mathrm{c}=4.3(1)$~K.
From transverse field $\mu$SR we obtain an in-plane penetration depth $\lambda_\mathrm{ab}(0)=223(2)$~nm for FeS. The temperature dependence of the corresponding superfluid density $\lambda_\mathrm{ab}^{-2}(T)$ indicates a fully gapped superconducting state and is consistent with a two gap s-wave model. Additionally, we find that the superconducting $T_\mathrm{c}$ of FeS continuously decreases for hydrostatic pressures up to 2.2~GPa.
\end{abstract}
\pacs{
74.25.-q,  
74.70.-b,  
74.62.Fj,  
76.75.+i  
 }

\maketitle

The structurally simplest Fe-based superconductor FeSe shows superconductivity (SC) below $T_\mathrm{c}=8$~K  \cite{Hsu2008} which can be increased in bulk specimens to 37 K by hydrostatic pressure  \cite{Medvedev2009,Margadonna2009} and by chemical modification to 45~K \cite{Guo2010,Krzton-Maziopa2011,Wang2011,Burrard-Lucas2013,Scheidt2012,Krzton-Maziopa2012}. The highest $T_\mathrm{c}$ in Fe-based SCs of $\approx$~100~K is reported for FeSe monolayers \cite{He2013,Ge2014}.
Density functional theory (DFT) calculations  \cite{Subedi2008} very early pointed out the electronic similarity of FeSe to tetragonal iron sulfide FeS (mackinawite). Attempts to synthesize superconducting FeS however have failed until Lai {\it et al.}  \cite{Lai2015} recently presented a low temperature hydrothermal synthesis route resulting in stoichiometric FeS with $T_\mathrm{c}$~$\approx 4.5-5$~K.
In Fe-based materials SC usually emerges in proximity to a competing magnetic phase \cite{Paglione2010,Kordyuk2012} and magnetic fluctuations are believed to be important for the Cooper pairing \cite{Mazin2008,Kuroki2008}. For pure FeSe however orbital nematic order \cite{Baek2014,Wang2015,Shimojima2014} has recently been suggested as the competing state.
On the other hand, magnetic order was found under pressure  \cite{Bendele2010,Jung2015,Terashima2015} with an ordering temperature $T_\mathrm{N}$  and antiferromagnetic spin fluctuations \cite{Imai2009} increasing together with $T_\mathrm{c}$ indicating a intimate coupling of the two ground states in the FeSe systems also.

To date, not much is known about the magnetic properties of superconducting FeS. Contradictory ground states are predicted for FeS. These range from a stripe antiferromagnetic order with a large magnetic moment of 2.7~$\mu_\mathrm{B}$  \cite{Kwon2011} to a non-magnetic metallic state  \cite{Devey2008}.
Subedi {\it et al.}  \cite{Subedi2008} pointed out the closeness of FeS to an itinerant SDW instability and its strong sensitivity to structural parameters.
Hence, DFT calculations produce a non-magnetic result for the relaxed crystal structure, but show a magnetic state with a progressively larger moment if the anion height is increased to the value observed by X-ray diffraction (XRD). Recently, Lin {\it et al.} \cite{Lin2015} observed a huge magnetoresistance in FeS at low temperatures and anomalies in the magnetoresistance and Hall effect data below 80~K. These features were tentatively attributed to multiband effects rather than interpreted magnetically.

In this Letter, we present muon spin rotation and relaxation ($\mu$SR) measurements on the magnetic and superconducting properties of polycrystalline FeS. It is found that bulk SC microscopically coexists with static, low moment disordered magnetism with a transition temperature $T_\mathrm{N}\approx 20$~K. This closely resembles the coexistence of magnetism and SC in bulk FeSe under a hydrostatic pressure of 0.9~GPa and suggests that magnetism might play a vital role for the appearance of SC in FeS.
From the temperature dependence of the superfluid density we infer that the superconducting order parameter is fully gapped and that it is best described by a two gap s-wave model. Magnetisation measurements under hydrostatic pressures yield that $T_\mathrm{c}$ decreases nearly linearly with $-1.6$~K/GPa until it reaches the base temperature (1.4~K) of the cryostat at 1.65~GPa.

Polycrystalline specimens with a $T_\mathrm{c} = 4.3(1)$~K of tetragonal FeS have been prepared by a similar hydrothermal synthesis method as recently described  \cite{Lai2015}. The structural characterization of our samples by XRD has been reported previously \cite{Pachmayr2016} and showed no detectable impurity phases.
Also $^{57}$Fe Moessbauer spectroscopy measurements shown in the Supplemental Material \cite{FeSsupplement} reveal no secondary phase within the detection limit of a few percent.
$\mu$SR measurements were performed at the Paul Scherrer Institut (Switzerland), using the general purpose spectrometer (GPS) as well as the low temperature facility (LTF) instruments.
The data were analysed with the free software package \textsc{musrfit} \cite{Suter2012}.

\begin{figure}[t]
\centering{
\includegraphics[width=0.92\columnwidth]{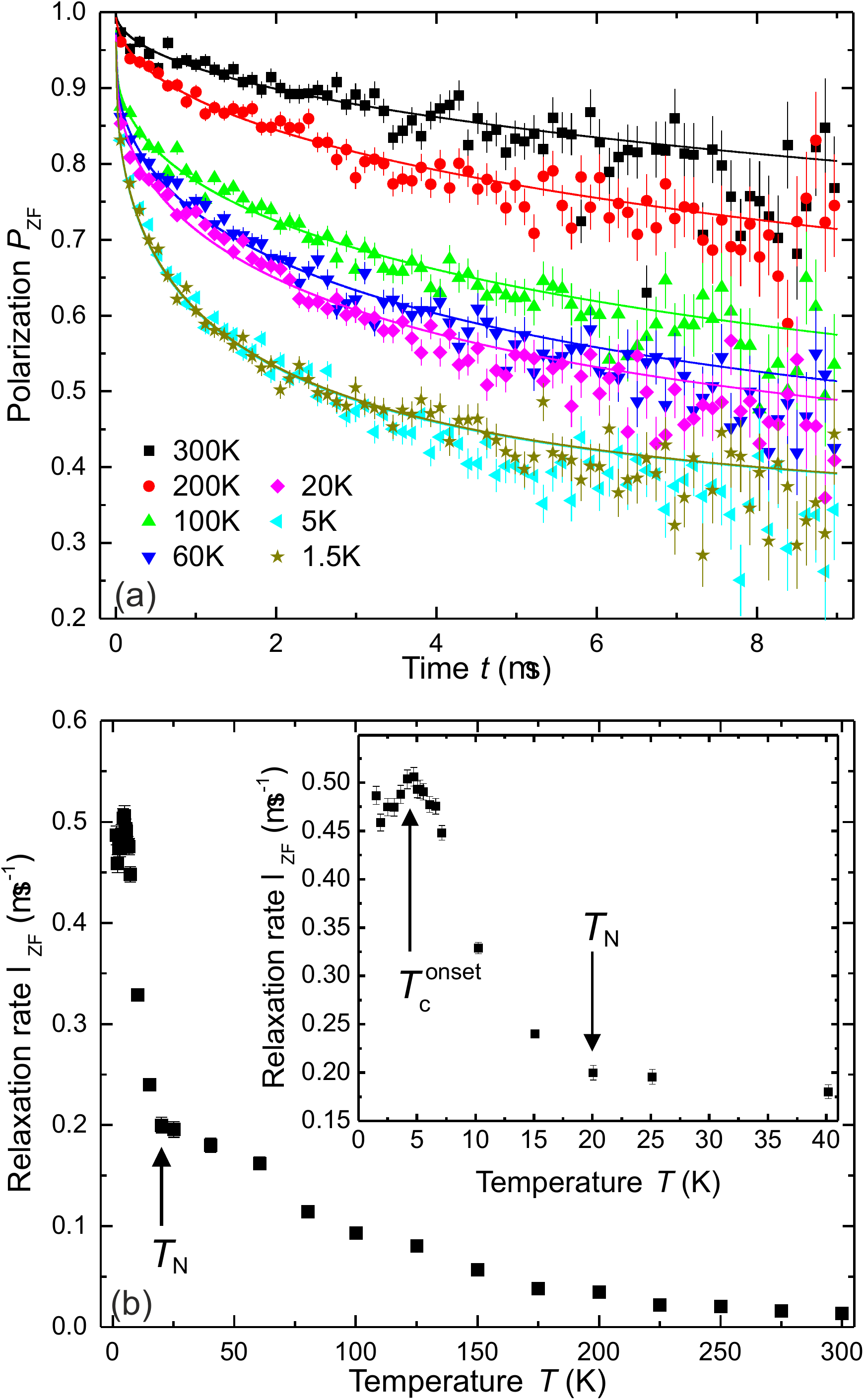}
 \caption{(a) Representative zero-field $\mu$SR spectra of FeS for various temperatures. (b) Zero-field relaxation rate $\lambda\mathrm{_{ZF}}$ as a function of temperature, showing the onset of magnetic order at $T\mathrm{_{N}} \approx \SI{20}{\kelvin}$.}\label{Fig:Fig1}
}
\end{figure}

Figure \ref{Fig:Fig1}(a) shows representative zero-field (ZF) spectra of FeS. At highest temperatures, a weak exponential relaxation is observed which is typical for a small amount of diluted ferromagnetic impurities with randomly oriented moments that create weak magnetic stray fields in the entire sample \cite{Walstedt1974}. Such a relaxation is commonly seen in Fe-based superconductors in the non-magnetic state e.g. in FeSe$_{1-x}$ \cite{Khasanov2008}.
Below \SI{200}{\kelvin}, a very fast loss of muon spin polarization at early times indicates strong internal fields in a minority volume fraction of the sample. This fraction increases with decreasing temperature to \SI{11}{\percent} at \SI{100}{\kelvin} and stays constant below \cite{FeSsupplement}. Even though the origin of this effect is not exactly known, we attribute this fraction to inclusions of small ferromagnetic impurities in our sample affecting a relatively large direct neighborhood via stray fields. These impurities are too small in volume to be detected by Moessbauer spectroscopy or XRD.
In the analysis of the data this has been taken into account by fitting the spectra with a sum of polarization functions for the minority and majority volume fractions ($f_\mathrm{maj}=89$\% below 100~K): $P\mathrm{_{ZF}}(t) = f\mathrm{_{maj}}P\mathrm{_{maj}}(t) + (1-f\mathrm{_{maj}})P\mathrm{_{min}}(t)$. The spectrum of the majority volume of our sample is best described by a root exponential relaxation:
\begin{equation}\label{Eq:ZF-spectra}
  P\mathrm{_{maj}}(t) = \frac{2}{3}\exp[-\sqrt{\lambda\mathrm{_{ZF}}t}] + \frac{1}{3}\,.
\end{equation}
The $2/3$ relaxing and $1/3$ non-relaxing components of the spectrum originate from the powder average of the internal fields with respect to the initial muon spin direction in our polycrystalline sample. The ZF relaxation rate $\lambda\mathrm{_{ZF}}$ of the majority phase signal is shown in Fig. \ref{Fig:Fig1}(b). It slowly increases below \SI{200}{\kelvin}, most probably due to increasing stray fields from the already magnetically ordered impurity phase, suggesting a fine mixture of the two phases. The sharp increase of $\lambda\mathrm{_{ZF}}$ below $T\mathrm{_{N}} \approx \SI{20}{\kelvin}$ indicates the onset of magnetic order in the full volume of the majority phase. The root-exponential form of the relaxation and the absence of coherent oscillations in $P(t)$ evidence a broad distribution of internal fields and therefore the short range/disordered nature of the magnetic state. Longitudinal field $\mu$SR experiments which are shown in the Supplemental Material \cite{FeSsupplement} prove that the magnetic order is static within the time window of the technique. The increase of $\lambda\mathrm{_{ZF}}$ stops at $T\mathrm{_{c}^{onset}} \approx \SI{4.4}{\kelvin}$ and $\lambda\mathrm{_{ZF}}$ decreases slightly below, see inset of Fig. \ref{Fig:Fig1}(b). This behavior indicates a microscopic coexistence and competition of the magnetic and superconducting order and has several times been observed in Fe-pnictide \cite{Wiesenmayer2011,Bernhard2012,Goltz2014b,Materne2015} and FeSe \cite{Bendele2010} superconductors by $\mu$SR. Note that the low temperature value of $\lambda\mathrm{_{ZF}}\approx 0.5$~$\mu$s$^{-1}$ is small indicating small internal fields of the order of $\lambda\mathrm{_{ZF}}/\gamma_\mu = 0.6$~mT (with $\gamma_\mu$ being the gyromagnetic ratio of the muon) pointing to very small values of the static moments.
From the internal magnetic field, the size of the Fe moments can be determined by $\mu$SR if the interstitial muon stopping site and the magnetic structure are known. If one assumes the same muon site \cite{Bendele2012} as in the isostructural FeSe and possible magnetic structures allowed by symmetry \cite{Bendele2012}, we can roughly estimate the magnetic moment in FeS to be in the order of $10^{-2}-10^{-3}\mu_\mathrm{B}$.
Our XRD measurements \cite{Pachmayr2016} on tetragonal FeS did not show a clear sign of an orthorhombic distortion which usually accompanies the SDW magnetism in Fe-pnictide superconductors. The orthorhombicity in these systems linearly increases with the ordered magnetic moment and follows a universal trend \cite{Jesche2008a,Wilson2010,Goltz2014b}. Assuming the same trend for FeS one obtains a possible orthorhombicity of maximal $4\times 10^{-5}$ which is below the resolution limit of our XRD and Moessbauer spectroscopy measurements which therefore could explain the absence of such a  signature in our corresponding data.


\begin{figure}[t]
  \centering
  \includegraphics[width=1.0\columnwidth]{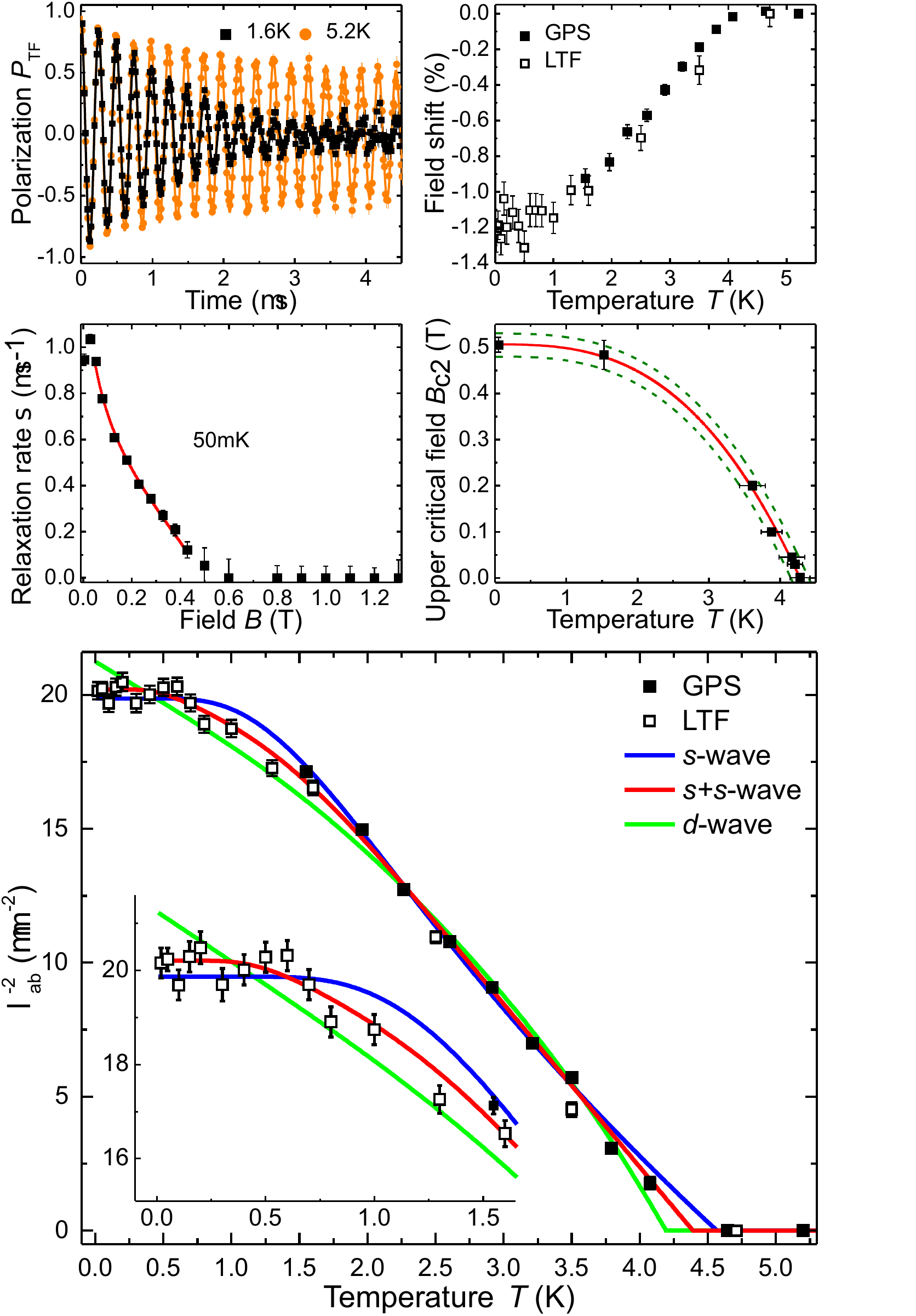}\\
  \caption{(a) Representative TF $\mu$SR spectra above and below $T_\mathrm{c}$ measured in 30~mT. (b) Diamagnetic field shift below $T_\mathrm{c}$. (c) Relaxation rate $\sigma_\mathrm{sc}$ as a function of field at \SI{50}{\milli\kelvin} (the red (solid) line is a fit using Eq. (\ref{Eq:sigmaOfB}). (d) Upper critical field $B_{\mathrm{c2}}$ obtained by $\mu$SR as a function of temperature. The red (solid) line is a phenomenological parametrization of $B_{\mathrm{c2}}(T)$. (e) Temperature dependence of $\lambda^{-2}_\mathrm{ab}$ calculated from the relaxation rate $\sigma\mathrm{_{sc}}(T)$. The solid lines are fits to the data using a single-gap \emph{s}-wave and \emph{d}-wave and a two-gap \emph{s}-wave model.}\label{Fig:Fig2}
\end{figure}
We now discuss the superconducting properties of FeS. We performed transverse-field (TF) $\mu$SR measurements with the external field applied perpendicular to the initial muon spin polarisation leading to a precession of the muon spin around the local magnetic field.
In a field cooled type-II superconductor, an additional Gaussian relaxation $\sigma\mathrm{_{sc}}$ of the $\mu$SR spectra appears below $T_\mathrm{c}$ due to the inhomogeneous field distribution of the flux line lattice (FLL) sensed by the muon ensemble. From $\sigma\mathrm{_{sc}}$, the absolute value of the magnetic penetration depth $\lambda$ can be obtained  \cite{Brandt1988,Brandt1988} providing a direct measurement of the superfluid density $n_\mathrm{s} \propto \lambda^{-2}$.   As an example, the TF-$\mu$SR spectra measured above and below $T_\mathrm{c}$ in a field of 30~mT are shown in Fig.~\ref{Fig:Fig2}(a).
The data were analyzed using the polarization function
\begin{equation}\label{Eq:TF-spectra}
  P\mathrm{_{TF}}(t) = \cos(\gamma\mathrm{_{\mu}}Bt + \phi)\exp[-\sqrt{\lambda\mathrm{_{TF}}t}-\frac{1}{2}(\sigma t)^2]\,,
\end{equation}
\noindent where $B$ is the mean field, $\phi$ the initial phase, $\lambda\mathrm{_{TF}}$ the root-exponential relaxation rate (in analogy to the ZF analysis) and $\sigma$ the Gaussian relaxation rate. The latter is a combination of a relaxation due to the FLL ($\sigma\mathrm{_{sc}}$) and a small temperature independent relaxation due to nuclear moments ($\sigma\mathrm{_{nuc}}$): $\sigma = \sqrt{\sigma^2_\mathrm{sc} + \sigma^2_\mathrm{nuc}}$. $\lambda\mathrm{_{TF}}$ and $\sigma\mathrm{_{nuc}}$ are fixed to their values above $T\mathrm{_{c}}$ for the analysis of the low temperature points. This is possible since the relaxation due to the magnetic order is relatively small and essentially temperature independent below 5~K as evidenced by the ZF raw data shown in Fig.~\ref{Fig:Fig1}.

The temperature dependence of $\sigma\mathrm{_{sc}}(T)$ and the diamagnetic field shift [see Fig.~\ref{Fig:Fig2}(b)] below $T\mathrm{_{c}}$ for a transverse field of \SI{30}{\milli\tesla} were measured down to 19~mK. Additional temperature scans were carried out at 45, 100 and \SI{200}{\milli\tesla} which allow one to extract the $T_\mathrm{c}$ as a function of external field that is depicted in Fig.~\ref{Fig:Fig2}(d). The two low temperature points were obtained by field scans [see Fig. \ref{Fig:Fig2}(c)] at fixed temperatures. In this case the upper critical field $B\mathrm{_{c2}}$ has been obtained by fitting \cite{Brandt2003}
\begin{eqnarray}\label{Eq:sigmaOfB}
  \sigma\mathrm{_{sc}}(b) = 0.172\frac{\gamma_\mathrm{\mu}\phi_{0}}{2\pi}(1-b)[1 + 1.21(1-\sqrt{b})^{3}]\lambda^{-2}_{\mathrm{eff}}
\end{eqnarray}
\noindent to the data, where $\lambda\mathrm{_{eff}}$ is the effective magnetic penetration depth, $b = B/B\mathrm{_{c2}}$ the reduced magnetic field, and $\phi_{0}$ the magnetic flux quantum. Since fitting $B\mathrm{_{c2}}$(T) with the Werthamer-Helfand-Hohenberg expression \cite{Werthamer1966} does not give satisfying results, we used the phenomenological parametrization $B\mathrm{_{c2}}(T) = B\mathrm{_{c2}}(0)[1-(T/T\mathrm{_{c}})^{\alpha}]$ with $T\mathrm{_{c}} = \SI{4.29(3)}{\kelvin}$, $\alpha = 2.8(2)$ and $B\mathrm{_{c2}}(0) = \SI{0.506(5)}{\tesla}$ to describe our data. The obtained $B\mathrm{_{c2}}(0)$ is in reasonable agreement with the value of $\approx \SI{0.4}{\tesla}$ reported by Lai {\it et al.} \cite{Lai2015}.
In plate-like single crystals of tetragonal FeS a strong anisotropy of the upper critical field of $B_{c2}^{\| ab}(0)/ B_{c2}^{\| c}(0) = 6-10$ has recently been reported  \cite{Borg2015,Lin2015}. In anisotropic polycrystalline superconducting samples the effective penetration depth $\lambda_\mathrm{eff}$ is dominated by the shorter penetration depth $\lambda_\mathrm{ab}$ and can be expressed by \cite{Fesenko1991} $\lambda_\mathrm{eff} = 3^{1/4} \lambda_\mathrm{ab}$.
Using this relation and Eq. (\ref{Eq:sigmaOfB}), the magnetic penetration depth $\lambda_\mathrm{ab}$ was calculated from the measured $\sigma\mathrm{_{sc}}(T)$ with the temperature dependence of $B_\mathrm{c2}$ as an input. The latter has only little influence on the absolute result for $\lambda_\mathrm{ab}$ which changes only by a few percent when the alternative $B\mathrm{_{c2}}(T)$ curves shown in Fig.~\ref{Fig:Fig2}(d) are used for the calculation, see \cite{FeSsupplement}.

Figure \ref{Fig:Fig2}(e) presents the temperature dependence of the superfluid density $n_\mathrm{s} \propto \lambda^{-2}_\mathrm{ab}$. Clearly $\lambda^{-2}_\mathrm{ab}(T)$ saturates towards low temperatures indicating a fully gapped superconducting state.
This observation is confirmed by fitting the data with a single- and two-gap BCS \emph{s}-wave as well as a single-gap \emph{d}-wave model \cite{Carrington2003,Prozorov2006}. Indeed, the two-gap \emph{s}-wave model gives the best result \cite{FeSsupplement} yielding $T_\mathrm{c} = \SI{4.33(3)}{\kelvin}$, $\lambda_\mathrm{ab} = \SI{223(2)}{\nano\meter}$, and the two superconducting gap values of $\Delta_{1}(0) = \SI{0.58(2)}{\milli\electronvolt}$ and $\Delta_{2}(0) = \SI{0.21(3)}{\milli\electronvolt}$ with a weighting factor $w = 0.83(4)$ for the contribution of the larger gap to the total superfluid density.
The gap to $T_\mathrm{c}$ ratios are $\Delta_{1}(0)/(k_\mathrm{B}T_\mathrm{c}) = 1.55(5)$ and $\Delta_{2}(0)/(k_\mathrm{B}T_\mathrm{c}) = 0.56(8)$ which is less than the BCS value of 1.76 showing that FeS is a weak coupling superconductor.
For completeness, we provide fundamental parameters that can be derived from our data which are a coherence length of $\xi = \SI{25.5(1)}{\nano\meter}$
(using the relation $B_\mathrm{c2} = \phi_{0}/(2\pi\xi^{2})$), a  Ginzburg-Landau parameter $\kappa  \approx 9$, and $B_\mathrm{c1} = \SI{7.19(2)}{\milli\tesla}$ (from $B_\mathrm{c1} = \phi_{0}/(4\pi\lambda^{2}_\mathrm{ab})\ln (\lambda_\mathrm{ab}/\xi)$ \cite{Tinkham1996}).

Since the superfluid density $\lambda^{-2}_\mathrm{ab}(T)$ determined by $\mu$SR is a quantity integrated over the whole Fermi surface, we can not reveal slight anisotropies in the gap function. Nevertheless, our data rule out a significant contribution of nodes in the superconducting gap. This is in contradiction to very recent thermal conductivity measurements \cite{Ying2015} on FeS foils which suggest $s$-wave SC with accidental nodes. Possibly both measurements could be conciliated by the fact that one of the gaps determined by $\mu$SR is extremely small. This two gap behaviour found in our FeS is very similar to the one observed in FeSe based superconductors investigated by $\mu$SR before, namely FeSe$_{0.85}$ \cite{Khasanov2008}, FeSe$_{0.97}$ \cite{Khasanov2010}, FeSe intercalated with different molecular spacer layers \cite{Burrard-Lucas2013,Biswas2013c}, and FeSe$_{0.5}$Te$_{0.5}$ \cite{Biswas2010}. In all these cases the presence of two $s$-wave gaps was deduced with one gap being a factor 3-7 smaller than the other and the contribution \emph{w} of the larger gap significantly larger than for the smaller gap.
There are also significant differences between FeS and FeSe. While FeSe is in the strong coupling limit, FeS appears to be a weak coupling superconductor. Also FeS possesses a much smaller $B_\mathrm{c2}$ (i.e. larger $\xi$) and a stronger anisotropy than FeSe \cite{Hsu2008,Lin2011,Her2015}.


\begin{figure}[t]
  \centering
  \includegraphics[width=0.92\columnwidth]{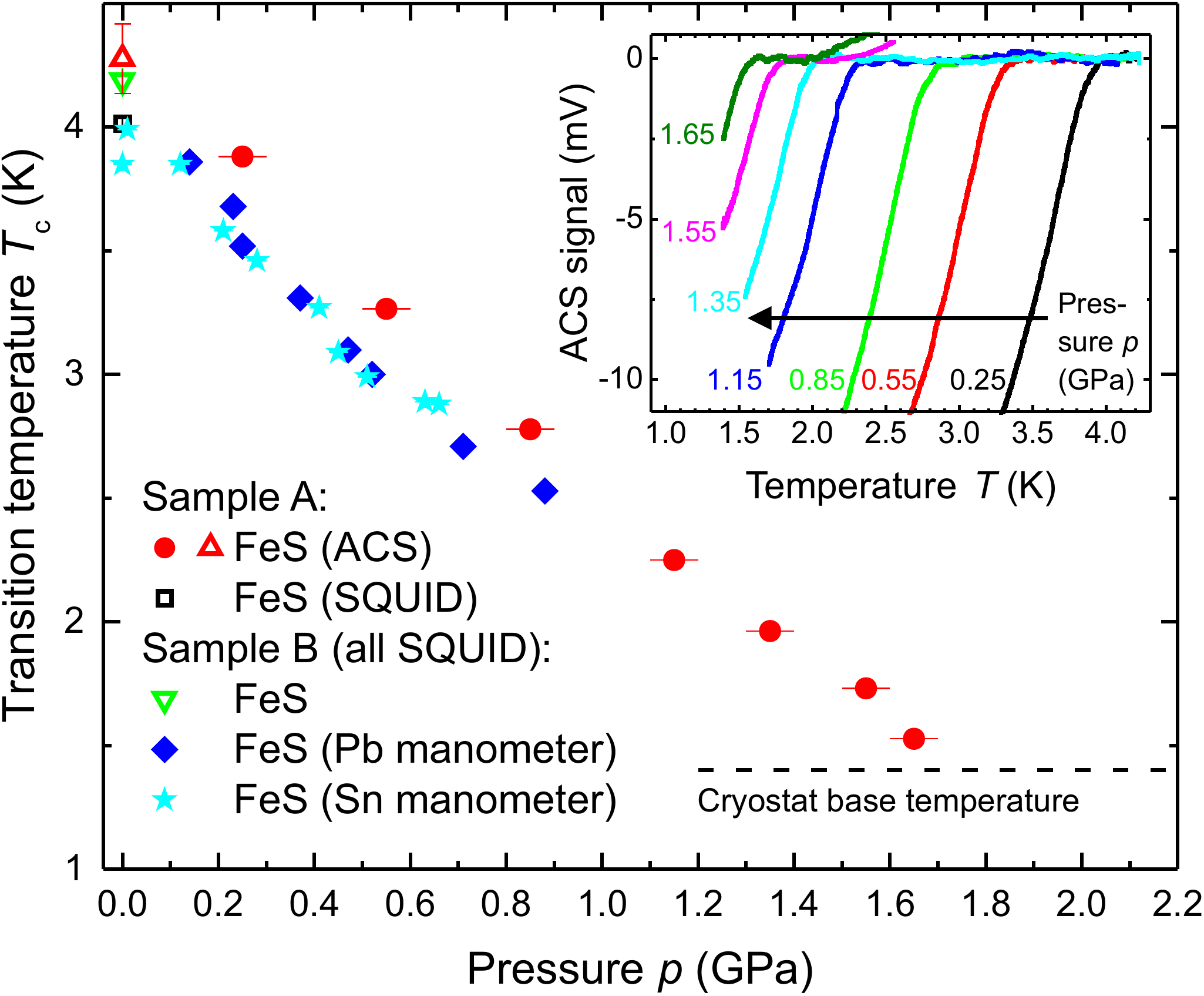}\\
  \caption{Pressure dependence of $T_\mathrm{c}$ for two different samples. Open symbols label measurements without a pressure cell. The low temperature pressure is either determined by Pb or Sn manometers or by subtracting a well known pressure loss from the room temperature value. Inset: ACS signal as a function of temperature for different pressures (linear background subtracted).}\label{Fig:Fig3}
\end{figure}

Motivated by the strong increase of $T_\mathrm{c}$ by hydrostatic pressure in FeSe \cite{Medvedev2009,Margadonna2009} we investigated the pressure dependence of $T_\mathrm{c}$ in FeS by SQUID and ACS magnetometry. Figure~\ref{Fig:Fig3} compiles the results for our measurements up to 2.2~GPa \cite{FeSsupplement}.
$T_\mathrm{c}$ decreases nearly linearly with pressure and reaches a value close to the base temperature of the cryostat at \SI{1.65(5)}{\giga\pascal} in qualitative agreement with a recent preliminary study \cite{Borg2015}. A measurement at \SI{2.20(5)}{\giga\pascal} did not show a transition between 1.4 and \SI{4.2}{\kelvin}.

Experimentally and theoretically it was found that the electronic properties of Fe-based superconductors are extremely sensitive to the local environment of Fe.
For example, an empirical relation of $T_\mathrm{c}$ and the anion height above the Fe layers shows that optimal $T_\mathrm{c}$s are reached for anion heights of 1.38~\AA  \cite{Mizuguchi2010} and regular (non-distorted) Fe-(pnictogen,chalcogen) tetrahedra \cite{Lee2008,Zhao2008,Lee2012}.
Our FeS powder samples \cite{Pachmayr2016} have nearly regular tetrahedra with Fe-S-Fe angles of 108.8(1)$^\circ$ and 109.8(1)$^\circ$ demonstrating that this parameter is not sufficient to induce high-$T_\mathrm{c}$s. Probably more importantly, it possesses an anion height of 1.32~\AA \, which is below the optimal value. Hydrostatic pressures probably further decreases the anion height as in FeSe \cite{Margadonna2009,Mizuguchi2010} which is in line with the observed decrease of $T_\mathrm{c}$. We would like to note however that $T_\mathrm{c}$ not necessarily needs to continue to decrease for higher pressures since several examples with a sudden reversal of this trend have been observed for FeSe \cite{Bendele2010,Sun2012,Izumi2015} and FeAs \cite{Tafti2013,Tafti2014,Tafti2015} based superconductors.

Ambient pressure FeS bears a remarkable reminiscence to FeSe at 0.9~GPa (in accordance with the smaller lattice constants of FeS). Both systems possess a magnetic transition at about 20~K and their $T_\mathrm{c}$s initially decrease as a function of pressure. In FeSe the suppression of $T_\mathrm{c}$ is related to a subtle competition of the magnetism with the superconducting state which for higher pressures transforms into a cooperative form of coexistence with higher $T_\mathrm{c}$s \cite{Bendele2010,Jung2015,Terashima2015}. Therefore, we suggest that further studies of the magnetic and superconducting properties of FeS under even higher pressures should be pursued to continue to investigate the commonalities and differences between the FeSe and FeS system to understand the existence/absence of high-$T_\mathrm{c}$s in the respective compounds.

In conclusion, we have shown that bulk SC in FeS microscopically coexists with low moment ($10^{-2}-10^{-3}\mu_\mathrm{B}$) disordered magnetism, that the superconducting order parameter is fully gapped and best described by a two gap s-wave model, and that $T_\mathrm{c}$ initially decreases as a function of hydrostatic pressure. These magnetic and superconducting properties strongly resemble those of FeSe at a pressure of 0.9~GPa in accordance with the smaller lattice constants in FeS and DFT calculations. However, there are also distinct differences to FeSe which are the weak coupling behavior and the larger coherence length.
Nevertheless, the bulk of our data suggest that the proximity to a magnetic instability might be crucial to SC in FeS and that higher $T_\mathrm{c}$s could possibly be reached by a proper chemical or physical modification of the structure in analogy to the case of FeSe.



We gratefully acknowledge the financial support of SH and ZG by the Swiss National Science Foundation (SNF-Grants 200021-159736 and 200021-149486) and of UP by the German Research Foundation (DFG-Grant Jo257/7-1). We further thank A. Schilling and R. Kremer for supporting the measurements of FeS under pressure.


%

\widetext
\clearpage
\begin{center}
\textbf{\large Supplemental Material:}

\textbf{\large Coexistence of low moment magnetism and superconductivity in tetragonal FeS and suppression of $T_\mathrm{c}$ under pressure}
\end{center}

\setcounter{equation}{0}
\setcounter{figure}{0}
\setcounter{table}{0}
\makeatletter
\renewcommand{\theequation}{S\arabic{equation}}
\renewcommand{\thefigure}{S\arabic{figure}}
\renewcommand{\bibnumfmt}[1]{[S#1]}
\renewcommand{\citenumfont}[1]{S#1}

\section{Moessbauer spectroscopy}

\begin{figure}[!b]
\includegraphics[width=0.88\columnwidth]{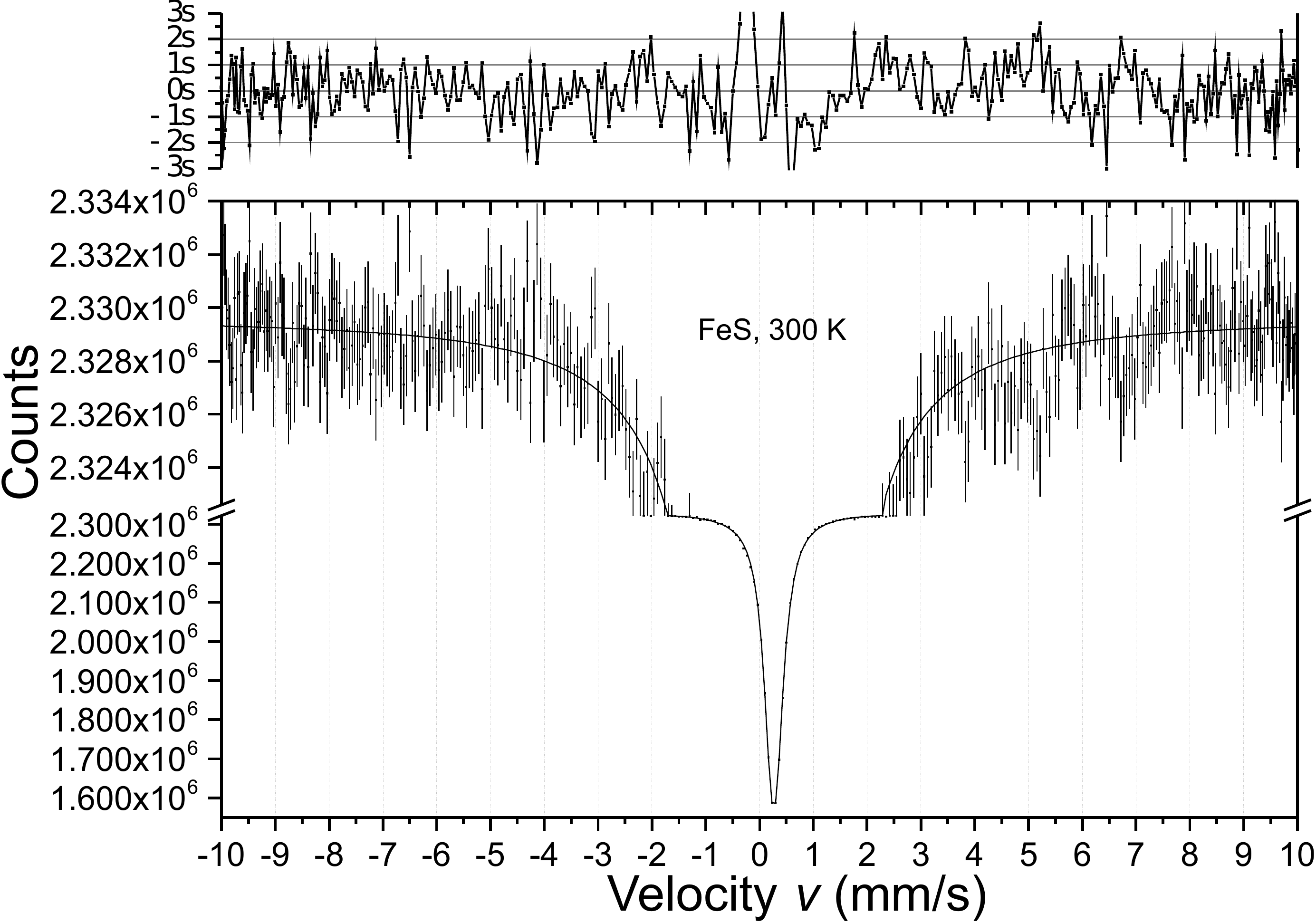}
\caption{The $^{57}$Fe-Moessbauer spectra prove the absence of common impurities. The fit is a doublet with a very small quadrupol splitting resulting in a single slightly broadened absorption line.}\label{MBS_RT_spectrum}
\end{figure}

$^{57}$Fe-Moessbauer spectroscopy was carried out to investigate the phase purity of the FeS sample. The room temperature measurement (Fig. \ref{MBS_RT_spectrum}) shows a doublet with a small quadrupolar splitting of $0.07(2)$mm/s indicative of the slightly distorted tetrahedral site in FeS. We can exclude ferromagnetic impurities such as elementary $\alpha$-Fe or Fe$_3$O$_4$ with a large Zeeman splitting and more than 1\,\% signal contribution. The same accounts for Fe$_2$S$_3$ or Fe$_3$S$_4$, even if it was nano-clustered \cite{Chang2008,Lyubutin2013} and consequently contributes only diffusely. It is unlikely, that the real volume fraction is strongly underestimated because the assumed impurities occur in well formed crystal structures, even if they were isolated nano particles, holding a high enough Debye temperature \cite{Yang2006} and thus large enough Debye-Waller factor, also at room temperature.

\newpage
\section{Muon spin rotation/relaxation spectroscopy (\muSR)}

\begin{figure}[!b]
  \centering
  \includegraphics[width=0.5\columnwidth]{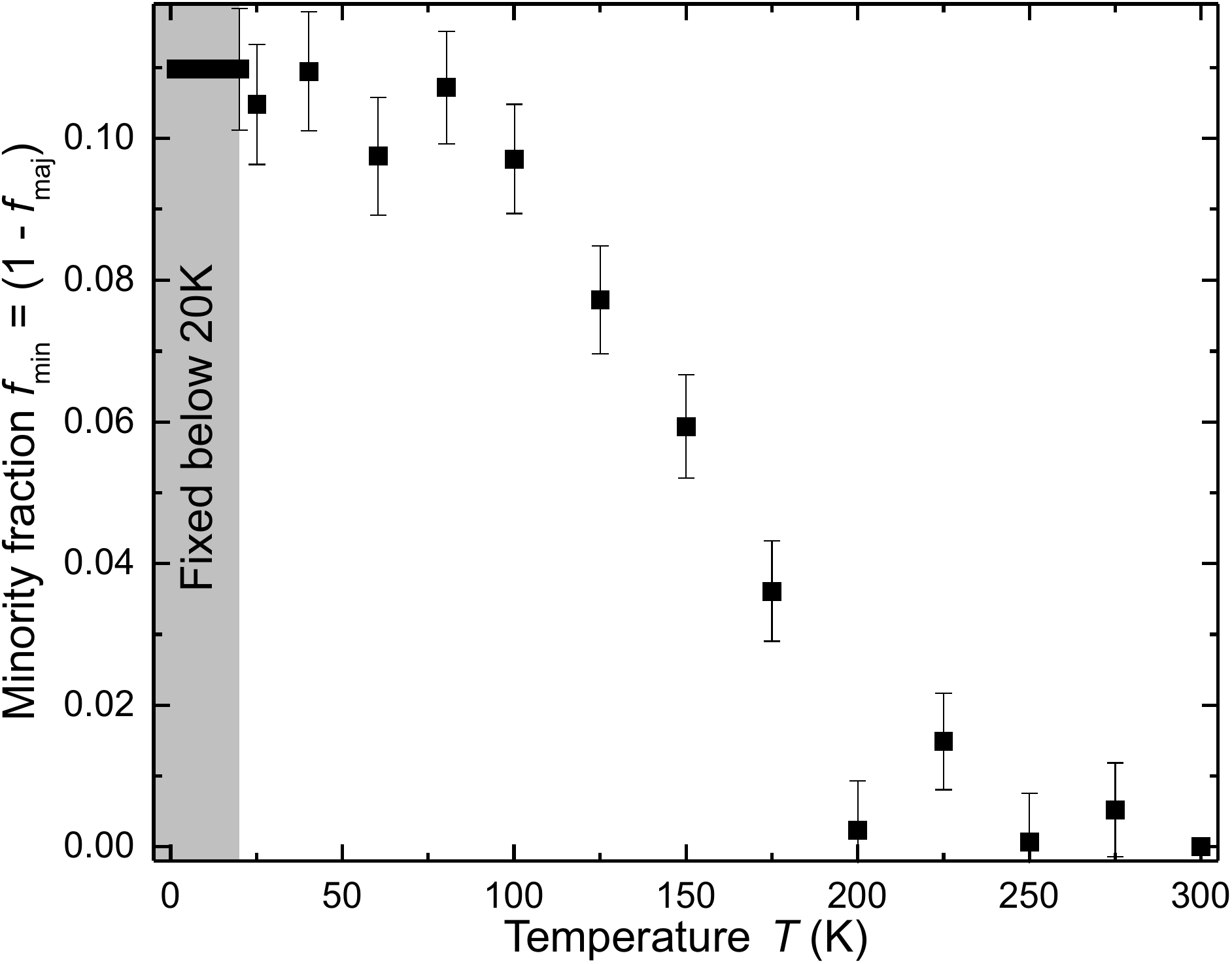}\\
  \caption{Evolution of the minority fraction $(f_\mathrm{min} = 1-f_\mathrm{maj})$ yielded by the fit to the ZF \muSR data of tetragonal FeS as a function of temperature.}\label{Fig:MinFrac}
\end{figure}

\begin{figure}[!b]
  \centering
  \includegraphics[width=0.5\columnwidth]{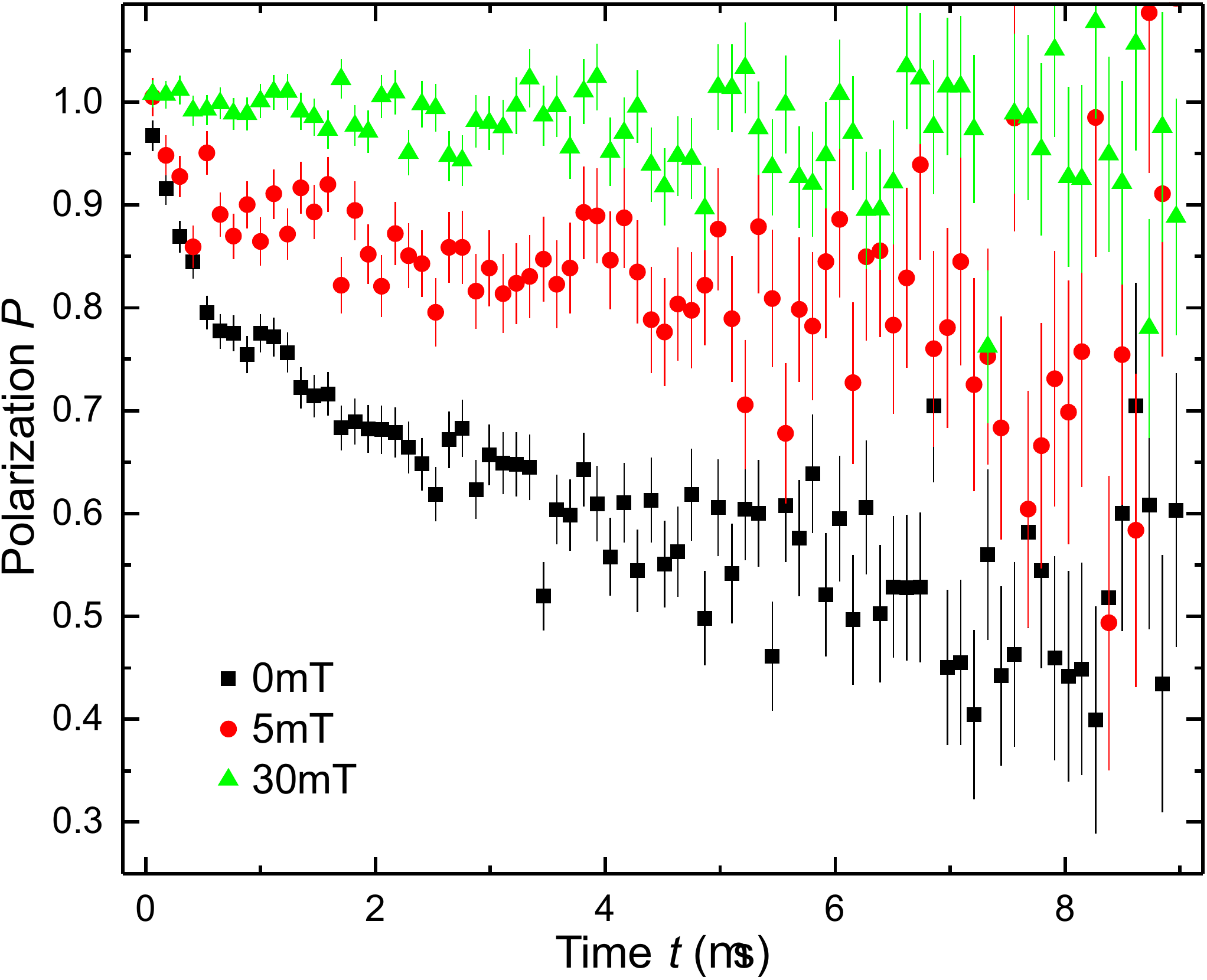}\\
  \caption{Longitudinal \muSR spectra of tetragonal FeS measured at \SI{5}{\kelvin}. A field of \SI{30}{\milli\tesla} is sufficient to completely decouple the muon spins from the internal field.}\label{Fig:decoupling}
\end{figure}

\begin{figure}[tbh]
  \centering
  \includegraphics[width=0.92\columnwidth]{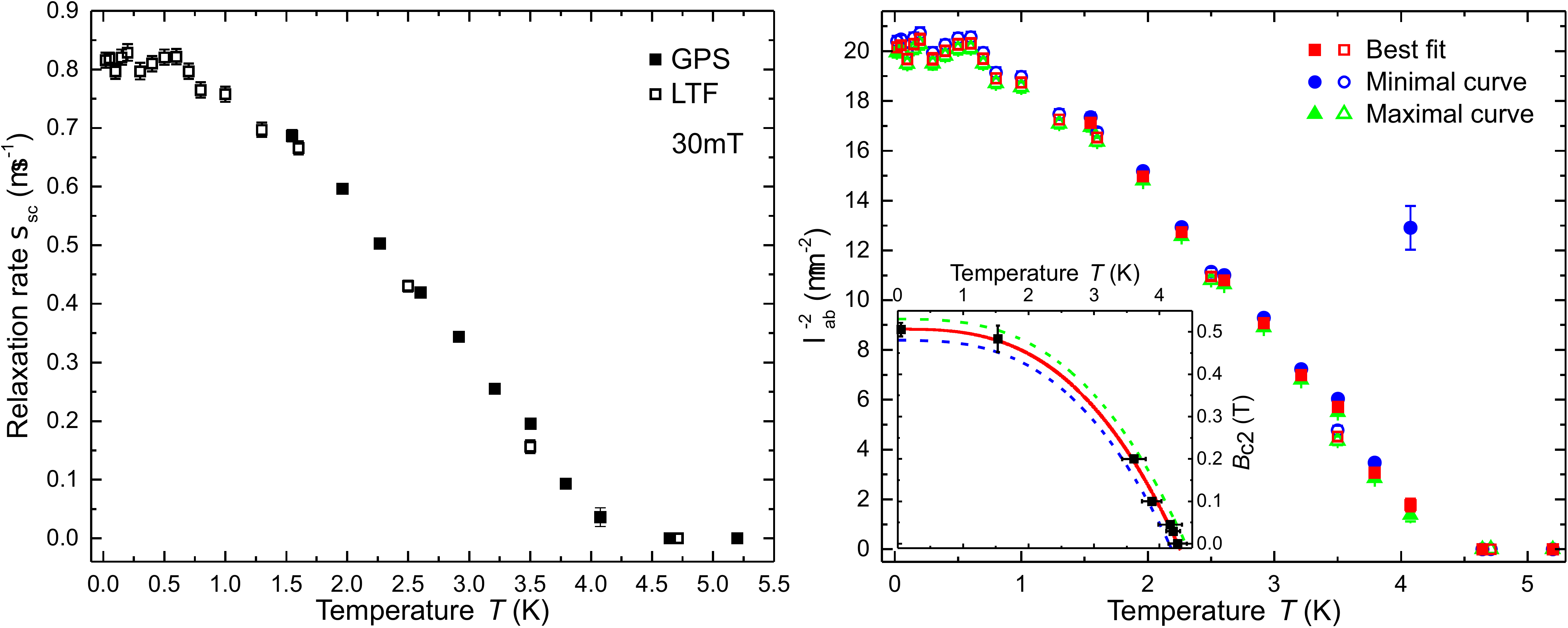}\\
  \caption{(a) Superconducting relaxation rate $\sigma_\mathrm{sc}$ measured in a transverse field of \SI{30}{\milli\tesla} as a function of temperature. (b) Temperature dependence of $\lambda^{-2}_\mathrm{ab}$ calculated from the relaxation rate $\sigma_\mathrm{sc}$, using the different temperature dependencies of the upper critical field $B_\mathrm{c2}$ shown in the inset. Full symbols denote measurements in GPS, open symbols denote measurements in LTF.}\label{Fig:SuperfluidDensity}
\end{figure}

$\mu$SR is based on the observation of the time evolution of the spin polarization $P(t)$ of an ensemble of initially polarized muons implanted in a sample \cite{Yaouanc2010,Reotier1997,Blundell1999,Amato2009}. The muon spins interact with the local magnetic field $B$ making it possible to extract information about the magnetic environment from $P(t)$. Zero field (ZF) $\mu$SR is an extremely sensitive magnetic probe which can detect magnetic moments as low as $10^{-4}$~$\mu_\mathrm{B}$. Due to the local character of the measurement the method is e.g. sensitive to short range and disordered magnetism as well as to the volume fractions in phase separated specimens. Figure \ref{Fig:MinFrac} shows the evolution of the minority fraction $f_\mathrm{min} = (1-f_\mathrm{maj})$ yielded by the fit of $P\mathrm{_{ZF}}(t) = f\mathrm{_{maj}}P\mathrm{_{maj}}(t) + (1-f\mathrm{_{maj}})P\mathrm{_{min}}(t)$ to the ZF \muSR data of tetragonal FeS ($P_\mathrm{maj}(t)$ given in the article, $P_\mathrm{min}(t) = \frac{2}{3}e^{-\lambda t} + \frac{1}{3}$). This fraction increases with decreasing temperature to maximally \SI{11}{\percent} at \SI{100}{\kelvin} and stays constant below. Therefore it was fixed to the \SI{20}{\kelvin} value for the analysis of the low temperature spectra. It is important to notice that this fraction represents a relatively large immediate neighborhood affected by magnetic stray fields from possible impurities, not the volume fraction of the impurities themselves. Further, the minority fraction relaxation rate is more than an order of magnitude larger than the one of the majority fraction, excluding a correlation between the two parameters.

Longitudinal field (LF) measurements, where an external field is applied parallel to the initial muon spin polarization direction, can be used to distinguish whether an observed depolarization originates from electronic fluctuations in the time window of \muSR or from a static magnetic field distribution \cite{Yaouanc2010}. In the latter case, an external field that is much stronger than the internal one ($B_\mathrm{ext}>10B_\mathrm{int}$) can "decouple" the muon spins from the static internal field, preventing a depolarization. As can be seen from Fig. \ref{Fig:decoupling}, a longitudinal field of \SI{30}{\milli\tesla} is enough to decouple the muon spins in our sample of tetragonal FeS, proving that the magnetic order observed by ZF measurements below \SI{20}{\kelvin} is static.

Transverse field measurements, where an external field is applied perpendicular to the initial muon spin polarization direction, were carried out to determine the superconducting properties of FeS. The additional Gaussian relaxation $\sigma_\mathrm{sc}$ of the \muSR spectra, which appears in a field cooled type-II superconductor due to the inhomogeneous field distribution of the flux line lattice (FLL) sensed by the muon ensemble, is shown in Fig. \ref{Fig:SuperfluidDensity}(a) as a function of temperature down to \SI{19}{\milli\kelvin} for a field of \SI{30}{\milli\tesla}. From this, the quantity $\lambda^{-2}_\mathrm{ab}(T)$ which is proportional to the superfluid density $n_\mathrm{s}$ was calculated, using Eq. (3) from the main article, the relation $\lambda_\mathrm{eff} = 3^{1/4}\lambda_\mathrm{ab}$ for anisotropic polycrystalline samples \cite{Fesenko1991} and the temperature dependence of the upper critical field $B_\mathrm{c2}$ [see inset of Fig. \ref{Fig:SuperfluidDensity}(b)]. Using the alternative $B_\mathrm{c2}(T)$ curves also shown in the inset of Fig. \ref{Fig:SuperfluidDensity}(b) only marginally changes the result, as can be seen in Fig. \ref{Fig:SuperfluidDensity}. Solely very close to the transition temperature $T_\mathrm{c}$, the value for $\lambda^{-2}_\mathrm{ab}(T)$ can diverge as it is the case for the minimal estimate of $B_\mathrm{c2}(T)$. A divergence of the superfluid density towards $T_\mathrm{c}$ is however unphysical and therefore indicates that the value of $B_\mathrm{c2}(T)$ is estimated too small.

The $\lambda^{-2}_\mathrm{ab}(T)$ data of FeS were fitted with a single- and two-gap BCS \emph{s}-wave as well as a \emph{d}-wave model using the following form for the superfluid density $\tilde{n}_\mathrm{s}$ normalised to its zero temperature value \cite{Carrington2003, Prozorov2006}:

\begin{eqnarray}\label{Eq:superfluidDensity}
  \tilde{n}_\mathrm{s}(T) = \frac{\lambda^{-2}(T)}{\lambda^{-2}(0)} &=& 1 + \frac{1}{\pi}\int_0^{2\pi}\int_{\Delta(\varphi,T)}^{\infty}\bigg(\frac{\partial f}{\partial E}\bigg)  \nonumber\\
  && \times \frac{E}{\sqrt{E^{2}-\Delta^{2}(\varphi,T)}}\mathrm{d}E\mathrm{d}\varphi\,,
\end{eqnarray}

\noindent where $f = (1 + \exp(E/k_\mathrm{B}T))^{-1}$ is the Fermi function. The superconducting gap was approximated by \cite{Carrington2003} $\Delta(\varphi,T) = \Delta(\varphi)\tanh[1.82(1.018(T_\mathrm{c}/T-1))^{0.51}]$, where $\Delta(\varphi) = \Delta_{0}$ for the \emph{s}-wave and $\Delta(\varphi) = \Delta_{0}\cos(2\varphi)$ for the \emph{d}-wave model. For the two-gap fit, the contributions of the two gaps were added together, weighted by a factor $w$ \cite{Carrington2003}: $\tilde{n}_\mathrm{s}(T) = w\tilde{n}_\mathrm{s1}(T) + (1-w)\tilde{n}_\mathrm{s2}(T)$. The two-gap \emph{s}-wave model gives the best fit, yielding the results given in Tab. \ref{Tab:gap-models}. For completeness, the results from the fits using the other models are included in the table.

\begin{table}[!tbp]
\begin{tabular}{lcccc}
  \hline
  \hline
  Model & Penetration depth & Transition temperature & Gap value & Gap ratio \\
  & $\lambda_\mathrm{ab}$ (\SI{}{\nano\meter}) & $T_\mathrm{c}$ (\SI{}{\kelvin}) & $\Delta(0)$ (\SI{}{\milli\electronvolt}) & $\Delta/(k_\mathrm{B}T_\mathrm{c})$ \\\hline
  \emph{s}-wave & 225(2) & 4.49(3) & 0.500(3) & 1.29(1) \\
  \emph{s}+\emph{s}-wave & 223(2) & 4.33(3) & 0.58(2), 0.21(3) & 1.55(5), 0.56(8) \\
  & & & with \emph{w} = 0.83(4) & \\
  \emph{d}-wave & 217(2) & 4.16(2) & 0.825(6) & 2.30(2) \\
  \hline
  \hline
\end{tabular}
  \caption{Parameters of the fits to the $\lambda^{-2}_\mathrm{ab}(T)$ data of FeS for different gap models.}\label{Tab:gap-models}
\end{table}


\section{Pressure dependence of the superconducting transition temperature $T_\mathrm{c}$}
Two different samples (A and B) originating from the same batch were measured using piston-cylinder type cells from MP35N and CuBe and 7373 Daphne oil as the pressure transmitting medium. The pressure was either determined in-situ at low temperatures using Pb and Sn manometers or at room temperature by subtracting the well known pressure losses at low temperatures. The transition temperature has been determined as the intersection of two linear fits to the data above and well below $T_\mathrm{c}$. In the case of the ACS measurements, a linear background from the pressure cell was subtracted before the determination of $T_\mathrm{c}$.



%

\end{document}